\documentclass[journal=jacsat,manuscript=article]{achemso}

\usepackage[version=3]{mhchem} 
\usepackage{color}
\usepackage{soul}
\usepackage{threeparttable}

\author{Shuai Liu}
\affiliation{Department of Electrical Engineering and Computer Science, The University of Michigan, Ann Arbor, Michigan 48109, USA.}
\email{shualiu@umich.edu}
\author{Yuheng Zhang}
\affiliation{Department of Electrical Engineering and Computer Science, The University of Michigan, Ann Arbor, Michigan 48109, USA.}
\author{Abdulkarim Hariri}
\affiliation{Department of Electrical Engineering and Computer Science, The University of Michigan, Ann Arbor, Michigan 48109, USA.}
\author{Abdur-Raheem Al-Hallak}
\affiliation{Department of Electrical Engineering and Computer Science, The University of Michigan, Ann Arbor, Michigan 48109, USA.}
\author{Zheshen Zhang}
\affiliation{Department of Electrical Engineering and Computer Science, The University of Michigan, Ann Arbor, Michigan 48109, USA.}
\email{zszh@umich.edu}

\title[An \textsf{achemso} demo]
  {Fabrication of Ultra-Low-Loss, Dispersion-Engineered Silicon Nitride Photonic Integrated Circuits via Silicon Hardmask Etching}

\abbreviations{IR,NMR,UV}

\begin{document}

\begin{tocentry}

\end{tocentry}

\begin{abstract}

Silicon nitride (Si$_3$N$_4$) photonic integrated circuits (PICs) have emerged as a versatile platform for a wide range of applications, such as nonlinear optics, narrow-linewidth lasers, and quantum photonics. While thin-film Si$_3$N$_4$ processes have been extensively developed, many nonlinear and quantum optics applications require the use of thick Si$_3$N$_4$ films with engineered dispersion, high mode confinement, and low optical loss. However, high tensile stress in thick Si$_3$N$_4$ films often leads to cracking, making the fabrication challenging to meet these requirements. In this work, we present a robust and reliable fabrication method for ultra-low-loss, dispersion-engineered Si$_3$N$_4$ PICs using amorphous silicon (a-Si) hardmask etching. This approach enables smooth etching of thick Si$_3$N$_4$ waveguides while ensuring long-term storage of crack-free Si$_3$N$_4$ wafers. We achieve intrinsic quality factors ($Q_i$) as high as $25.6 \times 10^6$, corresponding to a propagation loss of 1.6 dB/m. The introduction of a-Si hardmask etching and novel crack-isolation trenches offers notable advantages, including high etching selectivity, long-term wafer storage, high yield, and full compatibility with existing well-developed silicon-based semiconductor processes. We demonstrate frequency comb generation in the fabricated microring resonators, showcasing the platform's potential for applications in optical communication, nonlinear optics, metrology, and spectroscopy. This stable and efficient fabrication method offers high performance with significantly reduced fabrication complexity, representing a remarkable advancement toward mass production of Si$_3$N$_4$ PICs for a wide spectrum of applications.

\end{abstract}

\section{Introduction}

Silicon nitride (Si$_3$N$_4$) photonic integrated circuits (PICs) offer a unique combination of low optical loss, a large transparency window spanning from visible to mid-infrared, high power handling, high refractive index, moderate nonlinearity, and absence of free carrier absorption\cite{blumenthal2018silicon,ji2021methods}, enabling rapid advances across various fields, including quantum photonics\cite{arrazola2021quantum,chanana2022ultra,liu2024generation}, narrow-linewidth lasers\cite{jin2021hertz,xiang20233d}, frequency comb generation\cite{okawachi2011octave,gaeta2019photonic}, optical communication\cite{pfeifle2014coherent,marin2017microresonator}, among many others. In nonlinear optics, Si$_3$N$_4$ PICs are particularly attractive due to their ultra-low optical losses in tandem with appreciable Kerr nonlinearity, underpinning applications such as supercontinuum generation\cite{johnson2015octave}, parametric amplification\cite{ye2021overcoming}, and dissipative Kerr solitons (DKS)\cite{pfeiffer2017octave}. These applications rely heavily on tight optical confinement and precise dispersion engineering, which necessitates the use of thick Si$_3$N$_4$ layers with a typical thickness greater than 600 nm to enter the desired anomalous dispersion regime\cite{ji2024efficient,brasch2016photonic}.

However, growing thick Si$_3$N$_4$ films and fabricating high-quality PICs on top presents outstanding challenges. Low-pressure chemical vapor deposition (LPCVD) is commonly used to grow Si$_3$N$_4$ films with high quality and low propagation loss, but it induces substantial tensile stress that leads to cracking at film thicknesses exceeding 400 nm, severely limiting the performance and scalability of Si$_3$N$_4$ PICs \cite{luke2013overcoming}. To battle this issue, several approaches have been developed. For instance, the photonic Damascene process has successfully mitigated cracking and achieved optical losses ($\sim$ 1 dB/m) in Si$_3$N$_4$ waveguides by embedding them within a trench-like SiO$_2$ layer \cite{pfeiffer2018photonic,liu2021high}. Nevertheless, this process runs into challenges in maintaining precise waveguide dimension control, which is essential for applications demanding consistent mode dispersion and wafer-scale uniformity. Moreover, the photonic Damascene process involves multiple high-temperature, time-consuming steps and are hindered by issues such as the dishing effect \cite{ye2023foundry}, casting concerns over its cost effectiveness and complexity.

An alternative approach involves subtractive processing, which introduces cracking-isolation trenches to prevent crack propagation  within the Si$_3$N$_4$ films\cite{ji2021methods,ye2019high, xuan2016high}. Subtractive processing features a more uniform film thickness across the whole wafer by directly depositing the Si$_3$N$_4$ film onto patterned wafers, while also lends greater flexibility to fabricating large-scale patterns, such as arrayed waveguide gratings (AWG) or multimode interferometers (MMI), where the Damascene process could suffer from the dishing effect\cite{ye2023foundry}.  The subtractive processing based on e-beam lithography exposure and SiO$_2$ hardmask etching yielded the lowest optical losses to date--$Q_i = 37 \times 10^6$ and a propagation loss of 0.8 dB/m\cite{ji2017ultra}. However, in contrast to the Damascene process's SiO$_2$ trenches that undergo high-temperature thermal reflow to smooth sidewalls, the subtractive method requires highly optimized fabrication recipes and complex steps to prevent roughness accumulation and produce smooth Si$_3$N$_4$ waveguides\cite{ji2021methods}. More recently, deep ultraviolet (DUV) stepper photolithography was exploited in wafer-scale fabrication, achieving $Q_i$ values exceeding $28 \times 10^6$ \cite{ye2023foundry,ji2024efficient}. Although other approaches, such as multi-step slow LPCVD Si$_3$N$_4$ deposition with special rotation and annealing cycles \cite{el2019ultralow,el2018annealing}, sputtering-based Si$_3$N$_4$ film deposition\cite{zhang2024low}, and ICP-CVD growth using hydrogen-free precursors\cite{bose2024anneal,xie2022soliton}, have been developed to grow crack-free Si$_3$N$_4$ PICs, they have not yet yielded high \textit{Q} factors on par with those obtained with standard LPCVD Si$_3$N$_4$. Despite the aforementioned numerous advancements in developing thick Si$_3$N$_4$ films, to date a reliable and straightforward fabrication technique that preserves the high quality of Si$_3$N$_4$ PICs remains elusive.

In this work, we introduce an amorphous silicon (a-Si) hardmask etching technique for robust fabrication of ultra-low-loss, dispersion-engineered Si$_3$N$_4$ PICs with film thicknesses exceeding 800 nm. Our approach significantly simplifies the fabrication process while maintaining high performance, achieving an intrinsic quality factor as high as $Q_i = 25.6 \times 10^6$ with PECVD SiO$_2$ cladding, corresponding to a propagation loss of 1.6 dB/m . Furthermore, we introduce novel cracking-isolation trench designs along with the a-Si protective layer, allowing for long-term storage of crack-free Si$_3$N$_4$ wafers in a ready-to-use state. To showcase the prospect of the developed etching technique in nonlinear optics applications, we demonstrate efficient frequency comb generation in the fabricated microring resonators. Our work provides a highly robust, high-yield, and reliable method for producing high-quality Si$_3$N$_4$ PICs, particularly suited for foundry-scale manufacturing.

\section{Results and discussion}

\subsection{Fabrication flow: an overview}

\begin{figure}[h!]
\centering\includegraphics[width=14cm]{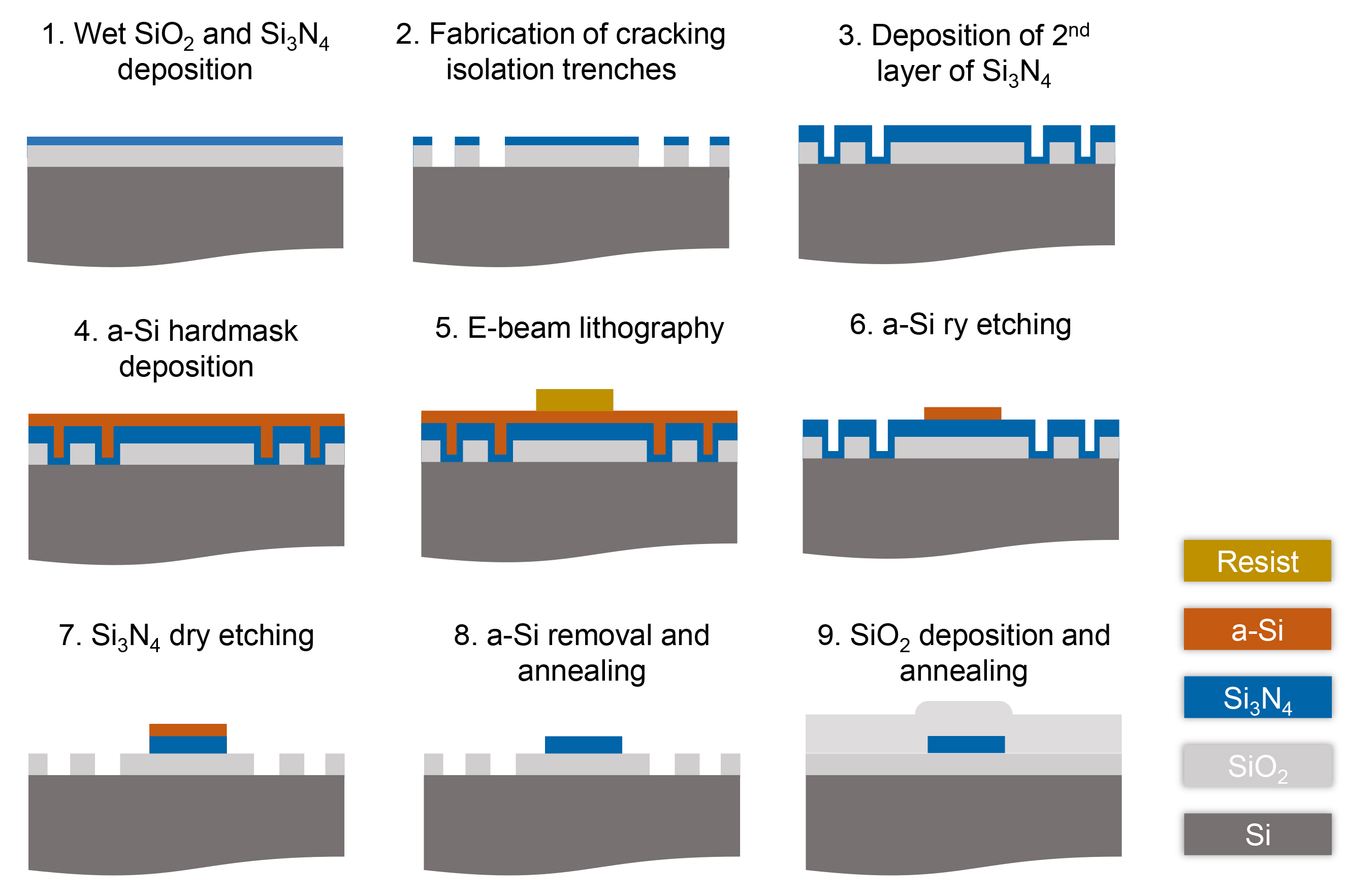}
\caption{{\bf Schematic of fabrication flow:} illustration of the amorphous silicon hardmask etching processes for ultra-low-loss Si$_3$N$_4$ photonic integrated circuits.}

\end{figure}

To grow crack-free, dispersion-engineered Si$_3$N$_4$-on-SiO$_2$-Si wafers, we employ a similar subtractive processing approach as that reported in Ref.\cite{luke2013overcoming, xuan2016high, ye2019high, ji2021methods, ye2023foundry}. Figure 1 depicts the fabrication flow conducted in the Lurie nanofabrication facility (LNF) at University of Michigan. Our process begins with the deposition of a 4-$\mu$m wet thermal SiO$_2$ layer on single-crystal silicon wafers. This is followed by the deposition of an initial thin Si$_3$N$_4$ layer ($\sim$ 380 nm) using low-pressure chemical vapor deposition (LPCVD). At this thickness, the tensile stress is manageable, minimizing the risk of Si$_3$N$_4$ film cracking. Next, UV photolithography is used to pattern the cracking isolation trenches, after which inductively coupled plasma reactive ion etching (ICP-RIE) is utilized to etch both the Si$_3$N$_4$ layer and the full thickness of the underlying SiO$_2$ layer. Following the etching process, we execute a thorough cleaning protocol, including oxygen plasma ashing, overnight Piranha cleaning, and two rounds of standard Radio Corporation of America (RCA) cleaning, to ensure that all organic residues, particles, and other contaminants are completely removed, such that any defects or cracking that could occur during the subsequent Si$_3$N$_4$ deposition are removed. A second round of LPCVD deposition of Si$_3$N$_4$ is then carried out to increase the total film thickness to about 800 nm, as required to enter the anomalous dispersion regime. Critical to our process is the elimination of high-temperature annealing that is usually carried out between the two LPCVD Si$_3$N$_4$ deposition rounds \cite{luke2013overcoming,ye2019high,ji2021methods} to mitigate stacked films at Si$_3$N$_4$ interfaces\cite{ji2024efficient} and reduce the wafer bowing issue. The deep trenches etched in the preceding steps effectively terminate the propagation of cracks formed along the wafer edges, resulting in a protected crack-free region. Immediately following the second round of Si$_3$N$_4$ growth, we deposit an LPCVD a-Si layer to serve as a hardmask. It is worth noting that while the trenches effectively prevent crack propagation during the LPCVD Si$_3$N$_4$ deposition, peeling-induced cracking may still occur in several weeks. In this regard, the introduced a-Si hardmask serves as a protective layer that prevents peeling and cracking formation during dicing or manual cleaving, enabling long-term storage of the deposited thick Si$_3$N$_4$ wafers and ensures that they can be readily used. In this study, both the 4-inch and 6-inch a-Si-protected Si$_3$N$_4$-on-SiO$_2$-Si wafers have been stored in a cleanroom environment for over 12 months, with no observed cracking in the protected regions. A direct comparison between wafers with and without the a-Si protective layer is provided in Section S3 of the Supporting Information. At this juncture, one may pattern waveguides using electron beam lithography (EBL) followed by Si$_3$N$_4$ RIE etching with the a-Si hardmask to create PICs. After a full removal of the a-Si hardmask, the etched Si$_3$N$_4$ chips undergo high-temperature annealing, followed by the PECVD SiO$_2$ cladding deposition and an additional annealing process. Further details of the fabrication steps can be found in Section S1 of the Supporting Information.

\subsection{Ready-to-use, crack-free Si$_3$N$_4$-on-SiO$_2$-Si wafers}

\begin{figure}[h!]
\centering\includegraphics[width=15cm]{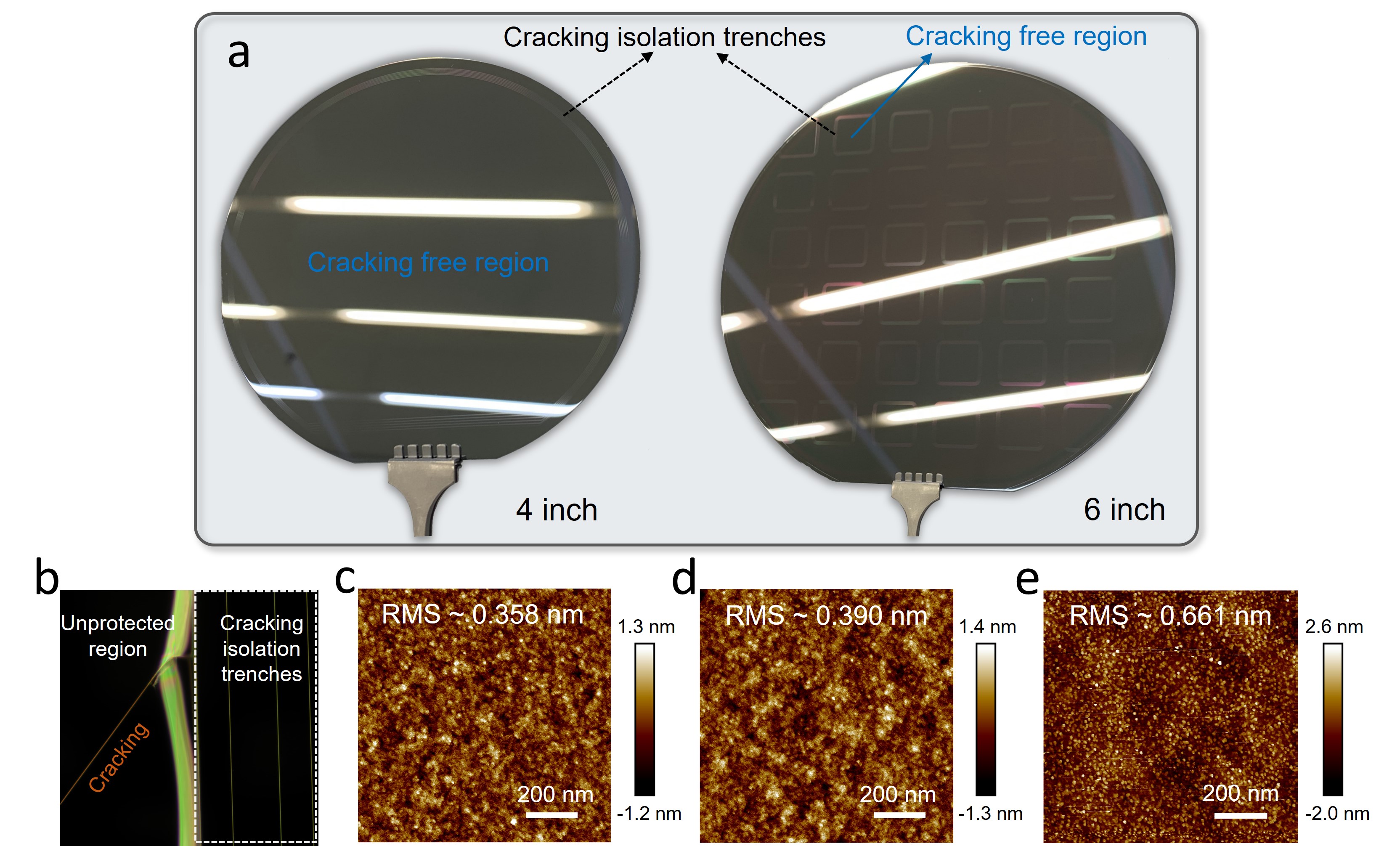}
\caption{ \textbf{Ready-to-use, crack-free Si$_3$N$_4$-on-SiO$_2$-Si wafers:} (a) Photograph of the fabricated 4-inch and 6-inch Si$_3$N$_4$-on-SiO$_2$-Si wafers with an 800-nm LPCVD Si$_3$N$_4$ and 700-nm a-Si hardmask. (b) Dark-field microscope image showing the effectiveness of the cracking isolation trenches in halting crack propagation. Atomic force microscopy (AFM) measurements of surface roughness: (c) the initial 380-nm Si$_3$N$_4$ film, (d) the 800-nm Si$_3$N$_4$ film after the second deposition, and (e) the a-Si hardmask layer, confirming the smoothness and high quality of the films at each stage of fabrication.   }
\end{figure}

Photographs of the fabricated Si$_3$N$_4$-on-SiO$_2$-Si wafers with the a-Si hardmask layer on top are presented in Fig. 2(a), in which two different designs of the cracking isolation trenches for the 4-inch (left) and 6-inch (right) wafers are displayed. Specifically, on the 4-inch wafer five 100 $\mu$m-wide cracking isolation trenches spaced 200 $\mu$m apart are introduced near the wafer edge. Crackings in a thick Si$_3$N$_4$ film typically originating at the wafer edges are effectively blocked by these trenches, as illustrated in the dark-field microscope image shown in Fig. 2(b). In contrast to the existing thick Si$_3$N$_4$ deposition approaches that are only capable of creating small crack-free regions confined to the central area of the wafer\cite{luke2013overcoming,ji2021methods,girardi2023superefficient}, our approach significantly enlarges the usable working area by virtue of the trenches at the wafer edge, as signified in the protected crack-free region marked in Fig. 2(a). Moreover, our new trench design not only improves yield but also ensures consistent crack-free performance. In this research, all of the 19 4-inch developed wafers remain intact within the protected crack-free regions. The crack-free design for the 6-inch wafer comprises a total of 32 crack-free dies, each containing six 80 $\mu$m-wide cracking isolation trenches spaced 120 $\mu$m apart (refer to Sections S2 and S3 of the Supporting Information for the design details). After the a-Si film deposition, the 6-inch wafers are diced into crack-free 2 cm $\times$ 2 cm dies prior to the subsequent EBL exposure. 

To assess the quality of the deposited films, we use atomic force microscopy (AFM) to measure the surface roughness at each deposition stage. The root mean square (RMS) surface roughness of deposited 380-nm LPCVD Si$_3$N$_4$ in the first round is measured to be 0.358 nm, as shown in Figure 2(c). The second round of deposition of LPCVD Si$_3$N$_4$ layer slighted increases the RMS roughness to 0.390 nm, as shown in Figure 2(d). Comparing both values with those reported in previous studies \cite{ji2017ultra} indicates high quality of the Si$_3$N$_4$ film. The exceptional surface smoothness is due to the a-Si hardmask, in lieu of poly-Si, deposited at a lower temperature to reduce e-beam or photon scattering during exposure, thereby minimizing additional resist edge roughness. Figure 2(e) shows that the RMS surface roughness of the a-Si film is 0.661 nm, which is sufficiently smooth for the subsequent EBL exposure.

\subsection{a-Si/Si$_3$N$_4$ hardmask dry etching}

\begin{figure}[h!]
\centering\includegraphics[width=13cm]{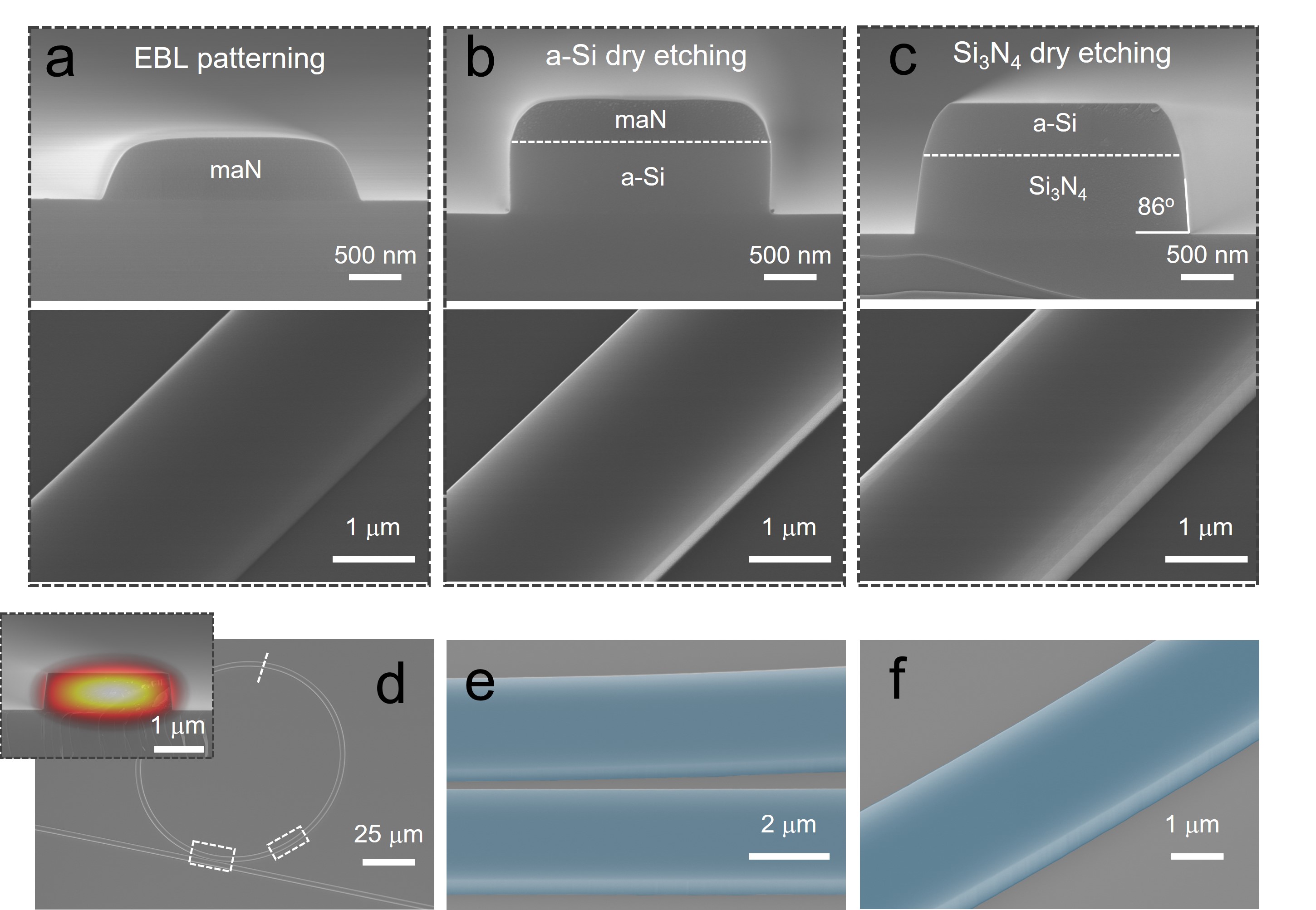}
\caption{\textbf{Characterization of a-Si/Si$_3$N$_4$ hardmask dry etching}: (a)-(c) Cross-sectional and tilted-view SEM images showing the e-beam resist, etched a-Si hardmask, and etched Si$_3$N$_4$ layers. (d) SEM image of the fabricated microring resonator after a-Si hardmask removal and annealing. The inset shows the corresponding cross-sectional view of the resonator overlaid with simulated quasi-TE waveguide modes. (e) and (f) False-color SEM images highlighting the smooth waveguide sidewalls in the bus waveguide-microring coupling region and along the microring waveguide, demonstrating smooth waveguide edges and sidewalls.}

\end{figure}

With the crack-free, 800-nm Si$_3$N$_4$-on-SiO$_2$-Si substrates with a-Si hardmask in hand, we optimize the core fabrication steps to produce ultra-low-loss Si$_3$N$_4$ PICs. The microring resonators are patterned using EBL exposure in a JEOL 6300 system. The maN 2405 resist is employed for single-pass writing at a uniform exposure dose of 700 $\mu$C/cm$^2$, with a beam current of 2 nA and a step size of 8 nm. After development, the maN 2405 resist is thermally reflowed on a hotplate to reduce edge roughness, as shown in the SEM images in Fig. 3(a). Subsequently, a two-step ICP-RIE etching process is developed to define the Si$_3$N$_4$ microring resonators. The maN 2405 patterns are first transferred to the a-Si hardmask layer, which is slightly over-etched using optimized ICP-RIE (LAM 9400) recipes with HBr and He gases. This process achieve a high etching selectivity of about 5:1 (a-Si: maN 2405), resulting in a vertical a-Si hardmask with smooth sidewalls, despite the maN 2405 resist having a slanted shape after reflow, as clearly depicted in Fig. 3(b). It should be noted that the a-Si hardmask is utilized due to its well-developed compatibility with industry silicon processing recipes and equipment, facilitating the smooth patterning and etching as compared to directly etching Si$_3$N$_4$. 

After removing the residual maN 2405 resist, the Si$_3$N$_4$ layer is etched using another ICP-RIE system (STS APS DGRIE), with a gas mixture of C$_4$F$_8$, CF$_4$, and He. In the process, C$_4$F$_8$ is introduced primarily due to its protective properties during Si etching, which, in tandem with Si$_3$N$_4$, ensures a high etching selectivity; CF$_4$ is added to adjust the C:F ratio as a knob that balances passivation and etching to simultaneously maintain high etching selectivity, high etching rate, and smooth sidewalls of the Si$_3$N$_4$ waveguides; He dilutes and stabilizes the plasma, leading to stable etching performance under low processing pressure. The etching process is optimized at proper levels of high RF power and low pressure to bias toward a physical-etching-dominated regime, achieving stable Si$_3$N$_4$ etching at high rates (approximately 340 nm/min) while being less prone to contamination in a shared ICP-RIE chamber environment caused by other etching processes such as aluminum metal hardmasks. As shown in Figure 3(a), the physical-etching-dominated process slightly erodes the corners of the a-Si hardmask. Nonetheless, benefiting from the high etching selectivity (Si$_3$N$_4$:a-Si $\sim$ 4:1), the process still yields vertical sidewalls with smooth surfaces at an angle of 86$^\circ$ for the Si$_3$N$_4$ waveguides. Over the past 12 months, consistent etching parameters have been maintained with rarely observed variations in etching rates, etching selectivity, sidewall angles, and sidewall roughness, indicating high reliability and stability of our optimized etching process. The fabrication process then proceeds with using XeF$_2$ etching (Xactix) to isotropically remove the residual a-Si, followed by standard RCA cleaning to eliminate the remaining particles. In the final step, the wafers are annealed at 1100°C in an N$_2$ environment for 6 hours. Fig. 3(d) shows the fabricated Si$_3$N$_4$ microring resonator after a-Si removal and annealing. The false-colored SEM images in Fig. 3(e) and 3(f) highlight the smooth edges and sidewalls at the waveguide-microring coupling region and along the microring waveguide, suggesting a low scattering loss and high-\textit{Q} factor.


\subsection{Si$_3$N$_4$ photonic integrated circuit characterization}

\begin{figure}[h!]
\centering\includegraphics[width=15.5cm]{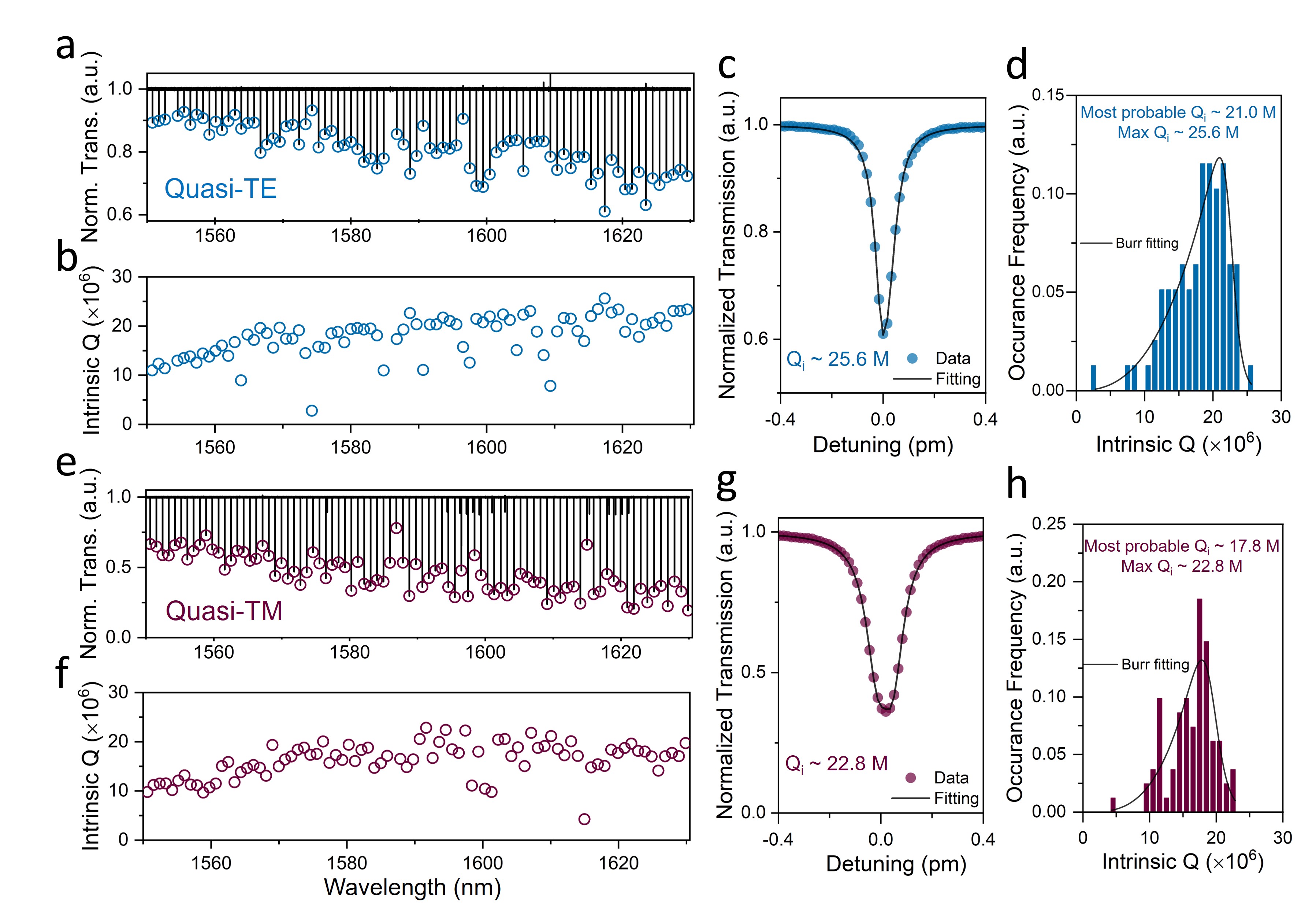}
\caption{ \textbf{Characterization of the fabricated Si$_3$N$_4$ microring resonators.} (a) Measured normalized transmission (Norm. Trans.) spectrum of quasi-TE waveguide modes from 1550 nm to 1630 nm. (b) Extracted intrinsic quality factors ($Q_i$) of the quasi-TE modes. (c) Zoom-in view of a resonance with an intrinsic linewidth of $\kappa_0 = 6.32 \times 10^{-2}$ pm, corresponding to $Q_i = 25.6 \times 10^6$ (million, M). (d) Histogram of the extracted $Q_i$ factors, showing a most probable value of $Q_i = 21.0$ M. (e) Normalized transmission spectra and (f) extracted $Q_i$ factors of quasi-TM waveguide modes. (g) Zoom-in view of a resonance with an intrinsic linewidth of $\kappa_0 = 7.17 \times 10^{-2}$ pm, corresponding to $Q_i = 22.9$ M. (h) Histogram of the $Q_i$ factors for the quasi-TM modes, showing a most probable value of $Q_i = 17.8$ M.}

\end{figure}

To characterize the optical loss of the fabricated Si$_3$N$_4$ PICs, we measure \textit{Q} factors of the microring resonators designed to operate in the anomalous dispersion regime. The microring resonator consists of a 2.8 $\mu$m-wide, 0.8 $\mu$m-high ring waveguide with a radius ($R$) of 200 $\mu$m. A straight bus waveguide with a width of 2.0 $\mu$m is coupled to the microring resonator with a coupling gap of 525 nm, enabling efficient excitation of the fundamental TE${_{00}}$ or TM${_{00}}$ mode. To achieve high fiber-to-chip coupling efficiency, we employ lens fibers in combination with tapered waveguide mode converters featuring a tip dimension of 0.2 $\mu$m $\times$ 0.8 $\mu$m, resulting in a coupling efficiency exceeding 70\%. The transmission spectra are obtained using a tunable laser (TSL-570) at low optical power to minimize thermo-optic effects and a low-noise photodetector (Newport 1811) to measure the transmitted light. The data are read out and recorded using a data acquisition card (DAQ). 

Figures 4(a) and (e) show the measured normalized transmission spectra for the quasi-TE and quasi-TM modes over the wavelength range of 1550 nm to 1630 nm, respectively. We then apply to these spectra a fitting model and analysis strategy similar to what Ref. \cite{pfeiffer2018ultra} adopted to extract $\kappa_0$ (intrinsic loss) and $\kappa_{\text{ex}}$ (coupling loss) and subsequently calculate the \textit{Q} factors using the formula $Q_{{i,\text{ex}}} = \lambda_0 / \kappa_{{0,\text{ex}}}$, where $\lambda_0$ is the resonant wavelength and $\kappa_{{0,\text{ex}}}$ represents the intrinsic and coupling linewidths. We assume in the analysis that all resonant modes operate in the under-coupled regime ($\kappa_{\text{ex}} < \kappa_0$). The extracted \textit{Q} factors for the quasi-TE${_{00}}$ and quasi-TM${_{00}}$ fundamental modes presented in Figs. 4(b) and (f) show that most resonances exhibit high $Q_i$ exceeding $10 \times 10^6$. 

Figure 4(c) shows a zoom-in view on the quasi-TE resonance at $\lambda_0$ = 1617.381 nm that exhibits the highest $Q_i = 25.6 \times 10^6$, extracted from a fitted intrinsic linewidth of $\kappa_0 = 6.32 \times 10^{-2}$ pm. The waveguide propagation loss $\alpha$ is then inferred as follows \cite{luke2013overcoming}: 

\begin{equation}
    \alpha = \frac{2\pi n_g}{Q_i\lambda_0},
\end{equation}

\noindent where $n_g$ is the group index, and determined to be $\alpha$ $\sim$ 1.6 dB/m. Similarly, the highest intrinsic quality factor for the quasi-TM resonances is inferred as $Q_i = 22.8 \times 10^6$, corresponding to a propagation loss of $\alpha$ $\sim$ 2.0 dB/m. A detailed description of the resonance fitting model and additional analysis is provided in Section S4 of the Supporting Information. Figure 4(d) and (h) present histograms of the extracted $Q_i$ factors. The distributions are modeled using Burr curves, and the maximum value of each fitted curve is defined as the most probable $Q_i$ factor \cite{pfeiffer2018ultra}, which amounts to $21.0 \times 10^6$ for the quasi-TE${_{00}}$ modes and $17.8 \times 10^6$ for the quasi-TM${_{00}}$ modes. These results underscore the consistent high-\textit{Q} performance endowed by the low propagation loss waveguides of the fabricated Si$_3$N$_4$ microring resonators.

Notably, a sharp drop in $Q_i$ from over $20 \times 10^6$ to approximately $10 \times 10^6$ in the wavelength range from 1580 nm to 1550 nm is observed for both the quasi-TE${_{00}}$ (Figure 4(b)) and quasi-TM${_{00}}$ (Figure 4(f)) modes. Beyond 1580 nm, the $Q_i$ values stabilize and drift around $20 \times 10^6$. This trend suggests that the lower \textit{Q} values near 1550 nm are not primarily attributed to fabrication-induced roughness, such as waveguide scattering loss, but are instead significantly influenced by material absorption--specifically, H-bond absorption in the Si$_3$N$_4$ waveguide and PECVD SiO$_2$ cladding. In Fig. S5 (Section S5 of the Supporting Information), we present additional transmission spectra of the quasi-TM modes across the 1490 nm to 1550 nm range for the same microring resonator shown in Figure 4. Another prominent degradation in the $Q_i$ factor is observed between 1550 nm and 1520 nm, with a minimum at around 1520 nm, where the H-bound absorption peak situates. Conversely, the $Q_i$ factor exhibits a gradual increase from 1520 nm to 1490 nm, raising $Q_i = 5.2 \times 10^6$ at 1520 nm to $Q_i = 13.0 \times 10^6$ at 1490 nm. However, since scattering loss is inversely proportional to $\lambda^4$\cite{gorodetsky2000rayleigh}, shorter wavelengths are anticipated to experience higher scattering loss compared to longer wavelengths. As such, the much higher $Q_i$ at shorter wavelength (1490 nm vs 1520 nm) in Fig.~S5 further corroborates that the relatively low observed $Q_i$ values are not predominantly caused by nanofabrication-induced scattering loss. Rather, they are due to the residual H-bonds that result in additional material absorption.

\begin{table}[h!]
    \centering
    \caption{Summary of state-of-the-art anomalous-dispersion LPCVD Si$_3$N$_4$ PICs and a comparison with the present work.}
    \label{tbl:fabrication_summary}
    \begin{tabular}{c c c c c c }
        \hline
        \textbf{} & \textbf{Method} & \textbf{Etch Mask} & \textbf{Cladding} & \textbf{Anneal Temp.} & \textbf{$Q_i$} \\
        & & & & \textbf{(°C)} & \textbf{($\times$ 10$^6$)} \\ \hline
        Ref\cite{ye2019high} & Subtractive & PR & LPCVD & 1100 & 14 \\
        Ref\cite{ji2017ultra} & Subtractive & SiO$_2$ & LPCVD & 1200 & 37 \\
        Ref\cite{ye2023foundry} & Subtractive & SiO$_2$ & LPCVD & 1200 & 14 \\
        Ref\cite{ji2024efficient} & Subtractive & SiO$_2$ & LPCVD & 1200 & 28 \\
        Ref\cite{liu2021high} & Damascene & SiN & LPCVD & 1200 & 30 \\
        Ref\cite{el2019ultralow} & Multi-step dep. & PR & HD-PECVD & 1200 & $\sim$ 10 \\ 
        \textbf{This work} & \textbf{Subtractive} & \textbf{a-Si} & \textbf{PECVD} & \textbf{1100} & \textbf{25.6}  \\ \hline
        
    \end{tabular}
    \vspace{0.5em}
    \begin{tablenotes}
        \small
        \item PR: Photoresist, HD-PECVD: High-density PECVD
    \end{tablenotes}
\end{table}

Despite the considerable potential in further reducing the optical loss by higher annealing temperature and LPCVD SiO$_2$ cladding, the demonstrated performance of the thick Si$_3$N$_4$ platform is already on par with state-of-the-art, as Table I summarizes. It is worth noting that the a-Si hardmask etching method offers benefits beyond ultra-low loss as discussed at the outset, including the ability for long-term storage in a ready-to-use state, ultra-high etching selectivity, high yield, robustness to RIE equipment variations, user-friendliness, and, most importantly, the ability to fully leverage well-developed equipment, processes, and infrastructure from the silicon industry.

\subsection{Frequency comb generation}

\begin{figure}[h!]
\centering\includegraphics[width=12cm]{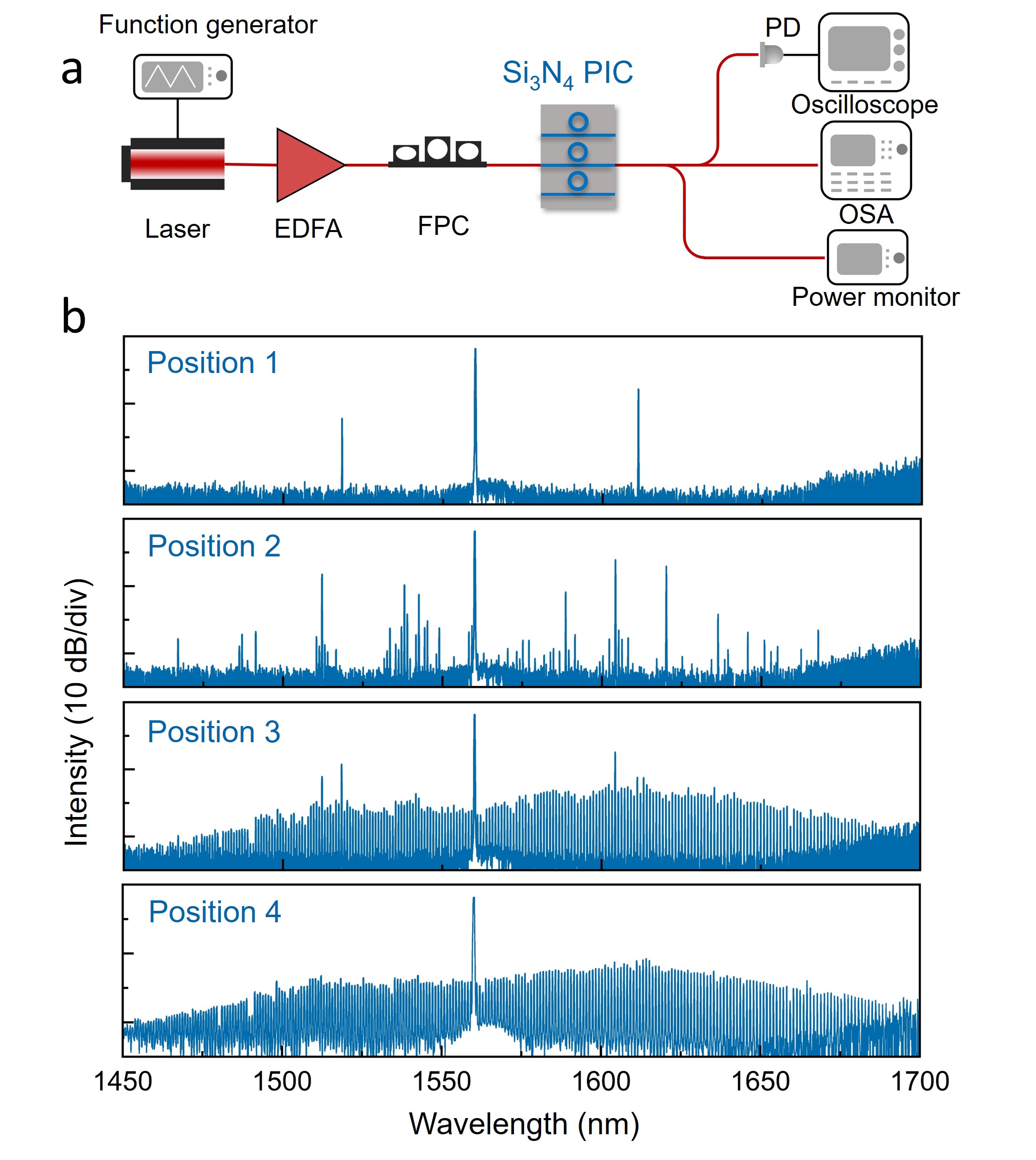}
\caption{ \textbf{Frequency comb generation in the Si$_3$N$_4$ microring resonators.} (a) Schematic of the experimental setup used for frequency comb generation. EDFA: erbium-doped fiber amplifier; FPC: fiber polarization controller; PD: photodetector; OSA: optical spectrum analyzer. (b) Evolution of the generated frequency combs as the pump wavelength is tuned closer to resonance, illustrated for positions 1 to 4, progressing from far detuned to near resonance.}
\end{figure}

We next demonstrate frequency comb generation to showcase our thick Si$_3$N$_4$ platform's prospect in nonlinear and quantum optics applications. Operating at an anomalous dispersion wavelength is crucial to generate frequency combs while low optical loss reduces the threshold and thereby the required pump power~\cite{kippenberg2011microresonator, brasch2016photonic}. The 800-nm thickness of the Si$_3$N$_4$ layer provides greater flexibility in attaining anomalous dispersion with a wider waveguide, while thinner Si$_3$N$_4$ films in general require narrower waveguides to achieve anomalous dispersion, which comes with increased optical loss due to a higher overlap between the waveguide modes and rough sidewalls. A more detailed discussion on engineering the waveguide dispersion can be found in Section S6 of the Supporting Information. 

In frequency comb generation, the same Si$_3$N$_4$ microring resonator geometry of a radius $R = 200\ \mu$m and waveguide dimension of 2.8 $\mu$m × 0.8 $\mu$m is employed in the experimental setup shown in Fig. 5(a): the output from a tunable pump laser is boosted by an erbium-doped fiber amplifier (EDFA) to about 50 mW. A fiber polarization controller (FPC) is used to fine-tune the polarization prior to injecting the light to excite the quasi-TE waveguide modes of the Si$_3$N$_4$ chip. Using a function generator, we control the pump wavelength of the tunable laser to approach a target high-\textit{Q} resonance near 1560 nm. The generated frequency combs displayed on an optical spectrum analyzer (OSA) is presented in Fig. 5(b), illustrating the evolution of the generated frequency comb as the pump wavelength is tuned into the resonance. Positions 1 to 4 in Fig. 5(b) show the frequency combs generated at different laser wavelengths in approaching the microring resonance. At position 1, initial four-wave mixing (FWM) sidebands are generated near the pump as its wavelength begins to approach resonance, marking the onset of comb formation. At position 2, increased cascaded FWMs and the generation of additional comb lines take place as more optical power is coupled into the resonator due to the pump wavelength getting closer to the resonance. At position 3, further cascaded FWM processes generate more comb teeth and create a denser frequency comb spectrum as the pump wavelength approaches even closer to the resonance. At Position 4, a broad chaotic modulation-instability (MI) frequency comb is generated, with each tooth spaced by one free spectral range (FSR), when the pump wavelength is very close to the resonance. The above results highlight the effectiveness of the fabricated Si$_3$N$_4$ microring resonators in generating dense and broadband frequency combs, demonstrating their potential in applications pertaining to optical communication, precision metrology, spectroscopy, among others. 

\section{Discussion}
The main focus of this research is to simplify the fabrication process using a-Si hardmask dry etching, while maintaining high performance. We have, however, identified additional techniques that can further reduce the optical loss substantially. Many of these techniques were developed in other high-\textit{Q} Si$_3$N$_4$ PICs studies and will be elaborated below:

1. Higher temperature annealing at 1200°C has been shown to effectively reduce H-bond absorption loss in both Si$_3$N$_4$ and SiO$_2$ layers\cite{henry1987low, liu2021high, li2013vertical, ye2023foundry, ji2024efficient}, crucial to achieve ultra-low loss near 1520 nm. 

2. LPCVD tetraethyl orthosilicate (TEOS) SiO$_2$ cladding followed by 1200°C annealing would address the gap-filling issues associated with PECVD SiO$_2$ cladding, thereby reducing additional scattering loss caused by the gap voids and at the same time minimizing H-bond absorption \cite{xuan2016high, pfeiffer2018photonic,liu2021high,ji2024efficient}.

3. Chemical mechanical polishing (CMP), as proven to significantly reduce surface scattering losses\cite{ji2017ultra}, would further smoothen the surface of the Si$_3$N$_4$ film. The RMS roughness of our Si$_3$N$_4$ film is $\sim$ 0.390 nm as shown in Fig. 2(c) and (d), while polished films demonstrated an RMS of $\sim$ 0.08 nm\cite{ji2017ultra}, highlighting the room in further reducing surface roughness. 

4. Optimized EBL exposure techniques, such as multi-pass writing, single-line smoothing, specialized fracturing designs, and post-exposure correction, would significantly reduce the edge roughness of the resist, which will in turn be carried over to the edge roughness of the etched Si$_3$N$_4$ waveguide\cite{bojko2011electron, ji2017ultra}.

5. Depositing an additional thin layer of Si$_3$N$_4$ can help recover defects on the waveguide sidewalls and surface introduced during RIE etching, thereby reducing scattering loss\cite{puckett2021422,sun2022low}.

6. Rapid thermal annealing can be exploited to repair broken Si-N bonds that are generated during exposure to UV light, electron beams, and plasma, due to their energies exceeding the bandgap of Si$_3$N$_4$. Such a damage typically occurs during the final fabrication steps, including photolithography alignment, e-beam evaporation, and RIE etching, during the fabrication of tapered waveguide couplers or deposition of on-chip metal microheaters and electrodes\cite{neutens2018mitigation,guo2024investigation, ji2024efficient}.

7. Optimizing the bus waveguide-to-microring resonator coupling geometry can enhance coupling fidelity\cite{pfeiffer2017coupling}, thereby reducing loss due to coupling to undesired waveguide modes. This optimization entails matching the dimensions of the coupling and microring waveguides and employing an adapted pulley coupling strategy to reduce the likelihood of high-\textit{Q} fundamental modes coupling to low-\textit{Q} higher-order modes \cite{spencer2014integrated, ji2021exploiting}. Moreover, engineered external strong coupling can be leveraged to further enhance the \textit{Q} factors \cite{liu2024formation}.

\section{Conclusions}
We have presented a novel fabrication process for ultra-low-loss, dispersion-engineered Si$_3$N$_4$ PICs using a-Si hardmask etching technique that addresses the challenges associated with fabricating thick Si$_3$N$_4$ films, including cracking due to tensile stress and long-term film stability. Our method gives rise to high intrinsic quality factors ($Q_i = 25.6 \times 10^6$) and low waveguide propagation loss ($\alpha$ $\sim$ 1.6 dB/m). Beyond the ultra-low-loss characteristics, the use of a-Si as a hardmask provides additional benefits, including robust long-term wafer storage, ultra-high etching selectivity, and scalability building on existing silicon industry infrastructure. Additionally, we have demonstrated frequency comb generation in our Si$_3$N$_4$ microring resonators, validating the platform's prospect in nonlinear and quantum optics and their pertinent applications. Overall, our approach offers a scalable, efficient solution for producing high-performance Si$_3$N$_4$ PICs, paving the way for their widespread adoption in nonlinear optics, quantum photonics, metrology, and beyond.


\section{Note}

The authors declare no competing financial interest.

\begin{acknowledgement}

The authors acknowledge the funding support from the National Science Foundation Grant No.~2326780, No.~2330310, and No.~2317471 and the University of Michigan.

\end{acknowledgement}

\begin{suppinfo}

The Supporting Information is available free of charge at:

The Supporting Information includes additional discussions: S1 Detailed fabrication flow; S2 Designs of the cracking isolation trenches; S3 Long-term storage of a-Si hardmask-protected Si$_3$N$_4$-on-SiO$_2$-Si substrates; S4 Resonance fitting models and additional examples; S5 Hydrogen absorption limited \textit{Q} factors; S6 Dispersion engineering of Si$_3$N$_4$ waveguide.

\end{suppinfo}

\bibliography{Main}

\end{document}